\documentclass[10pt,twocolumn,letterpaper,]{article}
\usepackage[T1]{fontenc}
\usepackage[utf8]{inputenc}
\usepackage{authblk}
\usepackage{cvpr}              
\usepackage{array} 
\usepackage{pifont} 
\usepackage[dvipsnames]{xcolor}

\definecolor{cvprblue}{rgb}{0.21,0.49,0.74}
\usepackage[pagebackref,breaklinks,colorlinks,citecolor=cvprblue]{hyperref}
\usepackage{multirow}
\usepackage[utf8]{inputenc}
\usepackage[margin=1in]{geometry} 
\usepackage{xcolor}
\usepackage{tcolorbox}
\usepackage{enumitem}

\definecolor{bg-gray}{RGB}{242, 242, 242}
\usepackage[table]{xcolor}
\usepackage{booktabs}   
\usepackage{multirow}   
\usepackage{amsmath}    
\usepackage[normalem]{ulem} 
\usepackage{booktabs}
\usepackage{multirow}
\usepackage{amsmath}
\usepackage{graphicx}  
\usepackage{float}     
\usepackage{graphicx}
\usepackage{flushend}
\usepackage{lipsum}
\usepackage{caption}
\usepackage{float}
\usepackage{amsmath}
\usepackage{balance}
\usepackage{algorithm}
\usepackage{algorithmic}
\usepackage[table]{xcolor}  
\usepackage{amssymb}      
\usepackage{booktabs}
\usepackage{multirow}
\usepackage{amsmath}
\usepackage[table]{xcolor}  
\usepackage{amssymb}      
\usepackage{booktabs}
\usepackage{multirow}
\usepackage{amsmath}
\usepackage{siunitx}      
\usepackage{hyperref}
\def\paperID{} 
\def\confName{CVPR}
\def\confYear{2026}

\author{
\vspace{-2em}
    Yusheng Dai$^{2,3}$, Zehua Chen$^{1,3\dag}$, Yuxuan Jiang$^{1,3}$,
    Baolong Gao$^{1,3}$, \\
    \vspace{-0.8em} Qiuhong Ke$^{2}$, Jianfei Cai$^{2}$, Jun Zhu$^{1,3\dag}$\\
    $^1$ Tsinghua University, Beijing, China  $^2$ Monash University, Melbourne, Australia \\ $^3$ Shengshu AI, Beijing, China 
}
\vspace{-2em}
\title{Omni2Sound: Towards Unified Video-Text-to-Audio Generation}

\begin{document}

\setlength{\abovecaptionskip}{5pt}
\setlength{\textfloatsep}{5pt}
\setlength{\intextsep}{-5pt}

\maketitle

\begin{abstract}
\vspace{-10pt}
\begingroup
\renewcommand\thefootnote{} 
\footnotetext{$^{\dag}$ Corresponding author.}
\endgroup
Training a unified model integrating video-to-audio (V2A), text-to-audio (T2A), and joint video-text-to-audio (VT2A) generation offers significant application flexibility, yet faces two unexplored foundational challenges: (1) the scarcity of high-quality audio captions with tight V-A-T alignment, leading to severe semantic conflict between multimodal conditions, and (2) cross-task and intra-task competition, manifesting as an adverse V2A-T2A performance trade-off and modality bias in the VT2A task. First, to address data scarcity, we introduce \textbf{SoundAtlas}, a large-scale dataset (470k pairs) that significantly outperforms existing benchmarks and even human experts in quality. Powered by a novel agentic pipeline, it integrates Vision-to-Language Compression to mitigate visual bias of MLLMs, a Junior-Senior Agent Handoff for a 5$\times$ cost reduction, and rigorous Post-hoc Filtering to ensure fidelity. Consequently, SoundAtlas delivers semantically rich and temporally detailed captions with tight V-A-T alignment. Second, we propose \textbf{Omni2Sound}, a unified VT2A diffusion model supporting flexible input modalities. To resolve the inherent cross-task and intra-task competition, we design a three-stage multi-task progressive training schedule that converts cross-task competition into joint optimization and mitigates modality bias in the VT2A task, maintaining both audio-visual alignment and off-screen audio generation faithfulness. Finally, we construct \textbf{VGGSound-Omni}, a comprehensive benchmark for unified evaluation, including challenging off-screen tracks. With a standard DiT backbone, Omni2Sound achieves unified SOTA performance across all three tasks within a single model, demonstrating strong generalization across benchmarks with heterogeneous input conditions.

\end{abstract}
\vspace{-15pt}
\section{Introduction}
\label{sec:introduction}

Early audio generation models typically rely on unimodal conditioning. Text-to-Audio (T2A) \cite{kreuk2022audiogen, liu2023audioldm, evans2024stableaudioopen, ghosal2023instruction, jiang2025controlaudio, jiang2025freeaudio} offers strong semantic fidelity and generalization but lacks dense temporal control. Conversely, Video-to-Audio (V2A) \cite{luo2023diff-foley, wang2024frieren, zhang2024foleycraft, chen2024videoguided} ensures fine-grained temporal synchronization with video, yet suffers from weak reasoning in complex scenes and unfaithful generation (e.g., unexpected music or speech) \cite{bain2025vintage, huang2025reasonaudio}. To address this, recent Video-Text-to-Audio (VT2A) methods \cite{chen2024hunyuanfoley, liu2025thinksound, cheng2025mmaudio, tian2025audiox,bain2025vintage} jointly condition on video and text. While VT2A achieves both strong semantic understanding and temporal alignment, its reliance on simultaneous inputs constrains its applicability \cite{tian2025audiox}. Crucially, most VT2A systems lack robustness \cite{liu2025thinksound,chen2024hunyuanfoley,bain2025vintage}, degrading sharply under missing-modality conditions (video-only or text-only).

These constraints motivate a unified framework natively supporting VT2A, V2A, and T2A within a single model. This unified paradigm aligns with the AIGC shift, eliminating the redundant architectures and deployment complexity of hard-switching between specialized models. Recent work has begun to advance this unified approach. MMAudio \cite{cheng2025mmaudio} introduces a multimodal joint training framework to improve V2A generation, optionally conditioning on text using large-scale text–audio pairs. Moreover, AudioX \cite{tian2025audiox} enhances flexibility by supporting broader modality combinations. Despite this progress, two challenges in the unified VT2A framework remain underexplored. \footnote{\href{https://omni2sound.github.io/}{https://omni2sound.github.io}}

\begin{figure}[!t]

    \centering
    \setlength{\belowcaptionskip}{0pt} 
    \includegraphics[width=1\columnwidth]{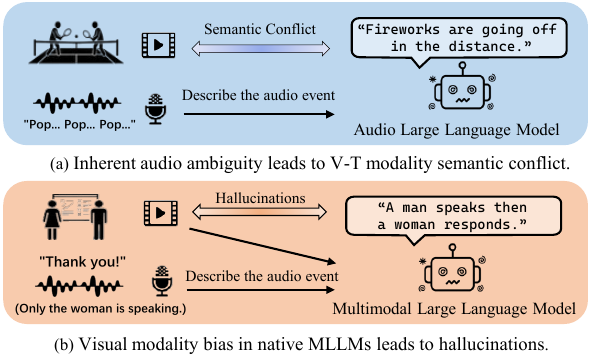}
\vspace{-10pt}
    \caption{Challenges in scaling high-quality audio captions.}
\label{fig:small}
\end{figure}

First, there is a scarcity of high-quality audio captions that are well-aligned with both audio and video cues. Most unified or specialized VT2A studies create their (V, T, A) training triplets by pairing videos (V) and their audio (A) with captions (T) generated solely from the audio \cite{tian2025audiox,chen2024hunyuanfoley}. However, this approach introduces severe semantic conflict in the multimodal training data (see Figure \ref{fig:small}): a frequent mismatch between the visual content and the (audio-only) text caption. This conflict is rooted in the audio modality's inherent ambiguity (e.g., a tennis hit vs. distant fireworks, or car engine noise vs. an electric drill). This fundamental ambiguity is then exacerbated by the limited capabilities of earlier audio-language models, which are prone to severe hallucinations (e.g., omissions and mislabels) \cite{chen2025detecting}. In our preliminary experiments, we found these modality conflicts caused by mismatches between V-T conditions directly lead to unstable convergence and a significant degradation in audio faithfulness. Unfortunately, there is still a lack of high-quality V-T-A triples for unified VT2A models training, as we further discuss in Section \ref{sec:related_datasets}.
    
Second, two critical types of task competition within unified VT2A frameworks remain underexplored. (1) Cross-Task Competition. Prior work, notably MMAudio \cite{cheng2025mmaudio}, established that incorporating T-A pairs enhances the generalization and quality of V2A generation. However, training a unified model to excel at both V2A and T2A presents a significant challenge: as shown in our preliminary experiment (Table \ref{tab:stage2_ablations}), this joint training introduces a severe T2A-V2A adverse trade-off, rooted in the heterogeneity between text and video modalities. Prioritizing one task during training consistently degrades the performance of the other, indicating a zero-sum optimization dynamic. (2) Intra-Task Competition. We also observe competition within the VT2A task itself. This competition manifests as a modality bias during generation process that undermines cross-conditional consistency, revealing two key failure modes: a bias towards text leads to poor A-V synchronization (Table \ref{tab:ablation_stages}), while a bias towards video exhibits low text-audio faithfulness in off-screen generation scenarios (Table \ref{tab:ablation_offscreen_track}).

To address data scarcity, we first introduce \textit{\textbf{SoundAtlas}} in Section \ref{sec:soundatlas_pipeline}, a large-scale, agent-generated multimodal audio-caption dataset. It augments the two largest audio datasets, VGGSound \cite{chen2020vggsound} and AudioSet \cite{gemmeke2017audio}, providing semantically rich and temporally detailed captions that even surpass human-expert quality (Table \ref{human_vs_machine_5col_new}). Built on current advanced multimodal foundation models (Gemini-2.5 Pro \cite{team2023gemini} and Qwen-2.5-VL \cite{qwen2025vl}), we develop a multi-turn, agentic annotation pipeline featuring a junior–senior agent handoff, vision-to-language compression, and post-hoc hallucination filtering. This pipeline delivers cost-controlled annotations while maintaining tight visual–audio–text (V–A–T) alignment and a markedly higher text-audio faithfulness than prior datasets. Interestingly, we find its quality is high enough to even correct human annotation errors in VGGSounder \cite{zverev2025vggsounder}.

Building on this dataset, we propose \textbf{\textit{Omni2Sound}} in Section \ref{sec:omni2sound_method}, a diffusion-based unified model supporting flexible input modalities while maintaining both audio-visual synchronization and high-fidelity generation. To address cross-task and intra-task competition, we introduce a three-stage progressive training schedule that departs from naive joint training. First, a \textit{large-scale T2A pretraining stage} establishes a robust generative prior, enabling minimal high-quality T2A replay in the subsequent stage to prevent catastrophic forgetting. Subsequently, our \textit{Multi-task Interleaved Training} integrates V2A and T2A tasks with high-quality VT2A triplets. Our central insight is that this VT2A data serves as a semantic bridge: by aligning the heterogeneous feature spaces of video and text, it effectively converts zero-sum cross-task competition into a cooperative optimization dynamic, thereby mitigating training resource contention. To resolve the intra-task competition, our third stage employs a \textit{decoupled Robustness Training}. We utilize two synergistic augmentations to balance cross-modal reliance: \textit{Text Dropout} penalizes text bias to enhance A-V synchronization, while \textit{Off-screen Synthesis} counteracts video bias to ensure textual faithfulness. This decoupled approach rectifies key failure modes, maintaining high-fidelity generation even in challenging, asymmetric input scenarios.

\begin{figure*}[t]
\centering
\includegraphics[width=2.1\columnwidth]{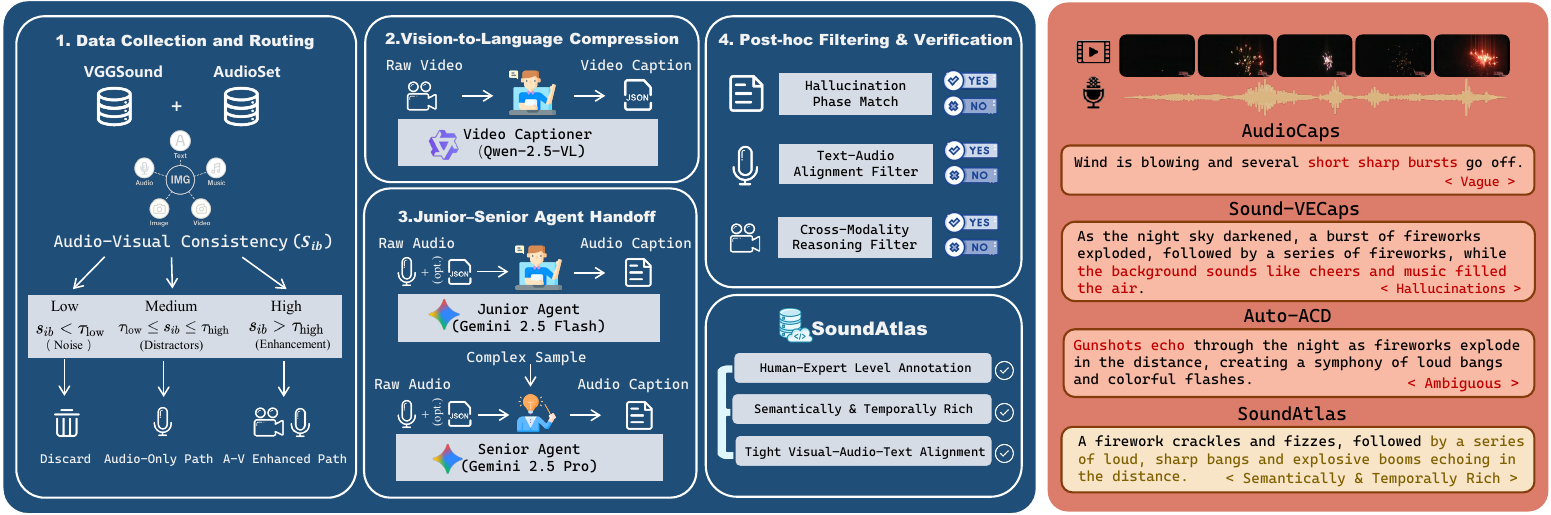}
\caption{Data Construction Pipeline of SoundAtlas (Left). Comparison against SOTA baselines and human annotations (Right) .}
\label{main_figure}
\end{figure*}

Finally, we construct \textit{\textbf{VGGSound-Omni}} in Section \ref{sec:benchmark}, the first comprehensive benchmark to establish a unified evaluation standard for VT2A, V2A, and T2A. It provides high-quality, human-verified annotations for all three tasks and introduces a challenging off-screen audio generation track. As a result, with a vanilla DiT \cite{Peebles2023DiT} backbone, Omni2Sound achieves unified state-of-the-art performance across all three tasks against both unified and specialized models, showing high-fidelity audio quality, tight audio-visual synchronization, and excellent generation faithfulness. 

\section{Related Works}
\label{sec:related_works}
\paragraph{Audio Caption Dataset.}
\label{sec:related_datasets}
Human-annotated benchmarks like \textit{AudioCaps} \cite{kim2019audiocaps} (46k) and \textit{Clotho} \cite{drossos2020clotho} (5k) offer high-quality alignment, but their limited scale, high cost, and lack of detail make them unsuitable for training modern, large-scale models. To address data scarcity, automated pipelines such as \textit{WavCaps} \cite{mei2023wavcaps} use LLMs to refine noisy web metadata (400k captions), and \textit{AudioSetCaps} \cite{lipping2024audiocaps} uses ALMs+LLMs to extract and aggregate details from audio, speech, and music, significantly increasing data volume. However, due to the inherent ambiguity of the audio modality, these audio-only methods suffer from high hallucination rates and can introduce cross-modal conflicts that destabilize VT2A training. Visually-enhanced (VE) annotation pipelines like \textit{Auto-ACD} \cite{han2024autoacd} and \textit{Sound-VECaps} \cite{yu2024soundvecaps} leverage visual cues as cross-modal constraints \cite{yang2025magic}. Despite their promise, existing VE pipelines adopt a separate-then-fuse design: unimodal models extract separate textual cues (e.g., image captions, audio tags), which are then merged by a final LLM. This is suboptimal, as the LLM operates on lossy textual representations rather than raw modalities, causing unimodal hallucinations to accumulate and amplify. Using native end-to-end multimodal models (e.g., \textit{Gemini} \cite{team2023gemini}) appears to be a natural alternative, yet as we show in Section~\ref{sec:soundatlas_pipeline}, this approach faces prohibitive costs and a pervasive visual bias that prevents audio-centric captioning. \textit{There remains a lack of a large-scale, high-quality visual–audio–text (V–A–T) aligned audio caption dataset suitable for training unified VT2A models.}
\paragraph{Unified Audio Generation Model.}

The audio generation paradigm is shifting towards unified, omni-modal frameworks \cite{xu2026tag}. This trend is initiated by \textit{MMAudio} \cite{cheng2025mmaudio}, which first integrates V2A and T2A but remains fundamentally V2A-centric, using T-A pairs merely as augmentation rather than optimizing T2A as a co-equal task. Subsequent works such as \textit{AudioX} \cite{tian2025audiox} expand the scope to more flexible modality combinations, while \textit{AudioGen-Omni} \cite{wang2025audiogenomni} and \textit{OmniSonic} \cite{pian2026omnisonic} further extend the paradigm to joint audio and speech generation. However, these approaches often rely on brute-force data scaling (e.g., \textit{AudioX} with over 9 million samples) without achieving commensurate performance gains. Critically, these early models \cite{cheng2025mmaudio, tian2025audiox, wang2025audiogenomni, pian2026omnisonic} largely overlook the inherent cross-task competition that arises from co-training diverse sub-tasks. \textit{UniFlow-Audio} \cite{xu2025uniflow} is the first to systematically address this issue by categorizing tasks into Time-Aligned (TA) and Non-Time-Aligned (NTA) classes and analyzing their competitive dynamics.  More recently, the trend has further extended toward joint audio and speech generation \cite{pian2026omnisonic,wang2025audiogenomni}. However, these analysis remains coarse-grained, without investigating the finer-grained competition within the TA category (i.e., V2A vs. T2A). Moreover, the challenging case of joint cross-modal generation (VT2A) remains unaddressed. \textit{Consequently, a fundamental study on task competitive dynamics within a unified VT2A framework remains absent.}

\section{SoundAtlas: V-A-T Data Construction}
\label{sec:soundatlas_pipeline}

Existing automated audio caption datasets often suffer from severe visual-audio-text (V-A-T) misalignment and high hallucination rates due to the limited capability of early ALMs \cite{lipping2024audiocaps, han2024autoacd, yu2024soundvecaps}. While recent native multimodal foundation models such as Gemini 2.5 \cite{team2023gemini, MMAR:2025, Qwen3Omni:2025, song2025hume} offer strong capabilities, we find that naively processing raw video-audio pairs is suboptimal for constructing audio caption datasets. Specifically, it incurs prohibitive costs (approx. \$10,275 per 1M samples; see Appendix~\ref{sec:cost_analysis}) and introduces inherent visual bias, where models hallucinate auditory labels for visually suggested but non-existent events, as shown in Figure~\ref{fig:small}.

To address these challenges, we introduce SoundAtlas, constructed through a cost-effective, multi-turn agentic annotation pipeline. As illustrated in Figure \ref{main_figure}, our pipeline integrates \textbf{vision-to-language compression} to mitigate visual bias, a \textbf{junior–senior agent handoff} to optimize cost-efficiency, and rigorous \textbf{post-hoc filtering} to ensure annotation fidelity. Full prompt instructions are provided in Appendix \ref{sec:prompt}.

\paragraph{A-V Consistency Routing.}
\vspace{-10pt}
We first apply A-V Consistency Routing to raw videos from AudioSet \cite{gemmeke2017audio} and VGGSound \cite{chen2020vggsound}. The key observation is that visual cues are reliable for high-consistency A-V clips but act as distractors in low-consistency ones, as shown in Figure \ref{main_figure}. We classify samples by their ImageBind alignment score ($s_{ib}$) using thresholds $\tau_{\text{low}}=0.20$ and $\tau_{\text{high}}=0.30$: (i) High-consistency samples ($s_{ib} > \tau_{\text{high}}$) enter the \textit{A-V Enhanced Path}; (ii) Medium-consistency samples ($\tau_{\text{low}} \leq s_{ib} \leq \tau_{\text{high}}$) are routed to the \textit{Audio-Only Path} to prevent visual hallucinations; and (iii) Noise ($s_{ib} < \tau_{\text{low}}$) is discarded.

\begin{table}[!t]
\centering

\caption{Semantic Faithfulness (CLAP Score) of Different Data Construct Pipelines on AudioSet and VGGSound.}
\label{tab:dataset_clap}
\setlength{\tabcolsep}{0.2mm} 
{\footnotesize 
\begin{tabular}{l cc cc}
\toprule
\multirow{2}{*}{\textbf{Method}} & \multicolumn{2}{c}{\textbf{AudioSet}} & \multicolumn{2}{c}{\textbf{VGGSound}} \\
\cmidrule(lr){2-3} \cmidrule(lr){4-5}
& \textbf{\scriptsize LA-CLAP $\uparrow$} & \textbf{\scriptsize MS-CLAP $\uparrow$} & \textbf{\scriptsize LA-CLAP $\uparrow$} & \textbf{\scriptsize MS-CLAP $\uparrow$} \\
\midrule
AudioSetCaps \cite{lipping2024audiocaps} & 0.330 & 0.397 & 0.351 & 0.421 \\
Sound-VECaps \cite{yu2024soundvecaps} & 0.370 & 0.425 & - & - \\
Auto-ACD \cite{han2024autoacd} & 0.396 & 0.437 & 0.409 & 0.457 \\
\textbf{SoundAtlas (Ours)} & \textbf{0.447} & \textbf{0.485} & \textbf{0.461} & \textbf{0.502} \\
\bottomrule
\end{tabular}
} 
\end{table}

\begin{table}[!t]
\centering
\caption{Caption quality comparison via MLLM-as-a-judge and human evaluation, reporting the Mean Win Rate for Semantic (MWR-S) and Temporal (MWR-T) alignment. Human-Expert refers to the human-annotated captions from AudioCaps \cite{kim2019audiocaps}.}

\label{human_vs_machine_5col_new}

\setlength{\tabcolsep}{0.5mm} 

{\footnotesize 
\begin{tabular}{l cccc} 
\toprule
\multirow{2}{*}{\textbf{Method}} & 
    \multicolumn{2}{c}{\textbf{MLLM Evaluation}} & 
    \multicolumn{2}{c}{\textbf{Human Evaluation}} \\
\cmidrule(lr){2-3} \cmidrule(lr){4-5}

& \textbf{MWR-S} $\uparrow$ & 
  \textbf{MWR-T} $\uparrow$ & 
  \textbf{MWR-S} $\uparrow$ & 
  \textbf{MWR-T} $\uparrow$ \\
\midrule
Auto-ACD \cite{han2024autoacd} & 0.39 & 0.41 & 0.31 & 0.26 \\
Human-Expert \cite{kim2019audiocaps} & 0.36 & 0.51 & 0.46 & 0.55 \\
\textbf{SoundAtlas (Ours)} & \textbf{0.75} & \textbf{0.58} & \textbf{0.71} & \textbf{0.69} \\
\bottomrule
\end{tabular}
} %
\end{table}





\paragraph{Vision-to-Language Compression.}
\vspace{-10pt}
This step implements our key insight: vision should be treated as a contextual constraint, not a primary input. We find that compressing the visual stream into a textual representation ($c_v$) effectively addresses both challenges identified above. First, it reduces cost by replacing the prohibitively expensive raw video input ($V+A$) with a cost-effective text-audio prompt ($c_v + A$). Second, it mitigates cross-modal hallucinations by removing direct visual bias and providing only semantic context (e.g., "A man and a woman are standing...") rather than the raw visual stream. Concretely, for samples $V$ routed to the \textit{A-V Enhanced Path}, we use Qwen-2.5-VL \cite{qwen2025vl} to analyze the video $V$ (without its audio $A$) and generate the textual representation $c_v = \text{Qwen}(V)$. For samples on the \textit{Audio-Only Path}, the visual context is set to null.


\paragraph{Junior–Senior Agent Handoff.}
\vspace{-10pt}
All samples then enter our handoff pipeline. Each sample is first assigned to the Junior agent, $G_{\text{junior}}$ (Gemini 2.5 Flash), which receives the audio $A$ and the optional visual context $c_v$, producing a caption $c_a = G_{\text{junior}}(A, c_v)$. This caption $c_a$ is flagged if it (i) triggers complexity criteria (text-based heuristics to identify complex audio scenes), (ii) contains high-frequency hallucination phrases, or (iii) fails a differentiated CLAP \cite{elizalde2023clap} check, $\text{CLAP}(c_a, A) < \tau_{clap}$, where $\tau_{clap}$ is $0.35$ for general audio and $0.15$ for music. Flagged samples are escalated to the Senior agent, $G_{\text{senior}}$ (Gemini 2.5 Pro). To control costs, the reasoning output of the Senior agent is limited to 128 tokens, yielding a more precise caption.
\paragraph{Post-hoc Filtering and Verification.}

Finally, all generated captions $c_a$ undergo two-stage verification. First, a CLAP (T-A) filtering model \cite{elizalde2023clap} ensures high Text-Audio faithfulness; captions where $\text{CLAP}(c_a, A) < \tau_{verify}$ are discarded. Second, for captions from the \textit{A-V Enhanced Path} ($c_v \neq \emptyset$), an V-A-T Verifier, $V_{\text{AVT}}$, checks that $c_a$ is a plausible acoustic description given $c_v$. Captions that pass all filters are accepted into the final dataset $\mathcal{D}_{\text{SoundAtlas (Ours)}}$, which augments VGGSound \cite{chen2020vggsound} and AudioSet \cite{gemmeke2017audio} with human-expert-level audio captions.






\subsection{Comparison with Existing Pipeline}
\label{sec:dataset_comparison}
We compare SoundAtlas against other automated pipelines~\cite{lipping2024audiocaps, han2024autoacd, yu2024soundvecaps} on high audio-visual consistency subsets sourced from AudioSet and VGGSound,  where ImageBind score $s_{ib} > 0.30$. 
As shown in Table~\ref{tab:dataset_clap}, SoundAtlas significantly outperforms all competitors on both LA-CLAP and MS-CLAP scores, demonstrating superior text-audio alignment. Additionally, we conduct a fine-grained MLLM-as-a-judge (Gemini 3.0 Pro \cite{team2023gemini}) evaluation on the intersection of AudioCaps and all compared datasets \cite{kim2019audiocaps}. As shown in Table~\ref{human_vs_machine_5col_new}, SoundAtlas achieves a substantially higher mean win rate in semantic alignment (MWR-S) and temporal alignment (MWR-T) than both the strongest baseline (Auto-ACD) and the Human-Expert annotations, across both semantic and temporal alignment. To mitigate potential evaluation bias, a follow-up human validation study is conducted, further corroborating our results (details in Appendix Section~\ref{sec:llm_judge_validation}). As illustrated in Figure~\ref{main_figure} (right), SoundAtlas demonstrates clear superiority over existing automated datasets, characterized by its richer semantic content and explicit temporal ordering.

\section{Omni2Sound: Unified VT2A Generation}

\label{sec:omni2sound_method}
Building on SoundAtlas, we propose Omni2Sound, a Diffusion-based unified VT2A model supporting collaborative (VT2A) and unimodal (V2A, T2A) control.

\subsection{Foundation Model Architecture}
\label{sec:foundation_model}

\begin{figure}[!t]
    \centering
    \setlength{\belowcaptionskip}{0pt} 

    \includegraphics[width=1\columnwidth]{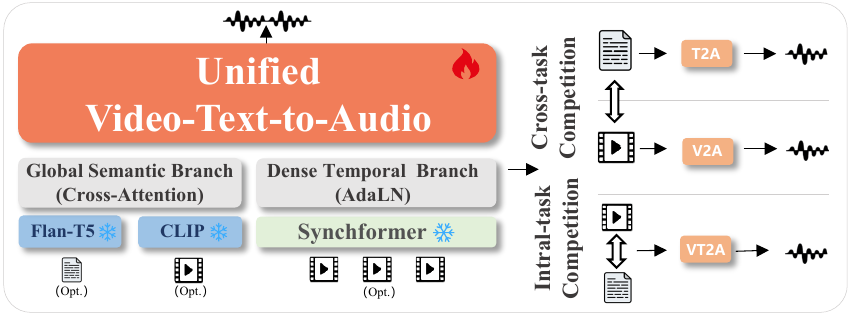}
    \caption{Overview of our unified VT2A framework, which integrates global semantics and temporal alignment, supporting flexible T2A, V2A, and VT2A generation.}
\label{fig:small1}
\end{figure}

We adhere to a principle of simplicity and scalability, adopting a standard Diffusion Transformer (DiT) backbone \cite{Peebles2023DiT} conditioned on latent features from a pre-trained audio VAE \cite{evans2024fast}. As shown in Figure \ref{fig:small1}, the backbone is conditioned on multimodal inputs using a decoupled injection approach, which is separated into two distinct branches: (1) \textbf{Semantic Branch (\textit{What})} and (2) \textbf{Temporal Branch (\textit{When})}. To capture global semantic context, we concatenate text embeddings from Flan-T5 \cite{chung2024scaling} ($F_t$) and visual features from CLIP \cite{radford2021learning} ($F_v$, sampled at 8 fps) along the temporal dimension, which are then injected via cross-attention layers.
Crucially, this design allows for flexible unimodal generation (V2A or T2A) by simply omitting the absent modality without requiring padding constraints. For the Temporal Branch, to ensure fine-grained synchronization, we follow \cite{cheng2025mmaudio} to utilize a Synchformer \cite{Viertola2024Synchformer} to extract dense visual-temporal features ($F_s$) and then inject it globally via Adaptive Layer Normalization (AdaLN).

This decoupled architecture effectively (1) achieves the flexibility of multi-condition frameworks like AudioX \cite{tian2025audiox}, supporting extensible conditions without architectural modification; and (2) maintains precise temporal alignment comparable to MMAudio \cite{cheng2025mmaudio} (powered by its well-designed MM-DiT architecture \cite{xu2024headrouter}).


\subsection{Three-stage Progressive Multi-task Training}
As established in Section \ref{sec:introduction}, naive joint training suffers from both cross-task and intra-task competition. To address these challenges, we design a three-stage progressive training schedule.

\paragraph{Stage 1: Large-scale T2A Pretraining.}
\vspace{-10pt}
We first pretrain the model on large-scale text-audio pairs without quality filtering. Following recent advances in latent generative modeling \cite{evans2024stableaudioopen, Peebles2023DiT, dai2025latentswap, wang2026geodesicnvs, xu2025b4m, xu2025context}, our DiT backbone (Section~\ref{sec:foundation_model}) learns to denoise a noisy latent $z_t$ at timestep $t$, conditioned on text embeddings $H_c$. The model $\epsilon_{\theta}$ is optimized with the standard L2 loss:
$$L = \mathbb{E}_{t,z_t,\epsilon} \|\epsilon - \epsilon_{\theta}(z_t, t, H_c)\|^2$$

This pretraining serves two purposes. First, it establishes a robust generative prior before introducing heterogeneous video conditions. Second, it enables a significantly lower T2A sampling frequency in the subsequent stage without catastrophic forgetting, thereby reducing resource contention.

\paragraph{Stage 2: Multi-task Interleaved Training.}
This stage addresses cross-task competition through interleaved task sampling. At each step, a single task $s \in\{V2A, T2A, VT2A\}$ is sampled from a categorical distribution $\text{Cat}(\pi)$, and a minibatch is drawn exclusively from its dataset $D_s$ for a single-task gradient update. This avoids within-batch loss mixing and stabilizes optimization. Our ablations (Table~\ref{tab:stage2_ablations}) reveal two key findings: (i) The VT2A task serves as a critical bridge that mitigates the adverse V2A-T2A trade-off, enabling their simultaneous optimization rather than zero-sum competition. (ii) With this bridge in place, a low T2A sampling frequency (e.g., $\pi_{T2A}=0.1$) on high-quality data suffices to prevent catastrophic forgetting. Together, these findings allow Stage 2 to focus primarily on video-conditioned tasks (V2A and VT2A), using T2A only sparingly to maintain its generative prior.

\paragraph{Stage 3: Intra-Task Resolution via Robustness Training.}
\label{sec:data_aug}
While Stage 2 resolves cross-task competition, intra-task competition (modality bias) persists, particularly for challenging scenarios such as off-screen generation. We therefore introduce a final robustness training stage. Crucially, this stage is decoupled from Stage 2: as shown in Table~\ref{tab:ablation_stages}, introducing robustness augmentations prematurely destabilizes the multi-task optimization, whereas applying them after convergence enhances cross-modal consistency without compromising generative quality.

This stage employs two complementary augmentations to enforce balanced reliance on both modalities: (i) \textit{Text Dropout.} We randomly drop tokens from the text prompt, creating ambiguity that compels the model to attend more to the visual stream. This counteracts text bias and strengthens audio-visual synchronization. (ii) \textit{Off-screen Synthesis.} We incorporate off-screen audio samples and augment the text prompt to describe them, producing training pairs where the audio content is absent from the video. This counteracts video bias and improves textual faithfulness for off-screen audio generation.

\begin{table*}[t]
\centering
\footnotesize
\setlength{\tabcolsep}{2.6mm}
\caption{Comparison on VGGSound-Omni benchmark: Omni2Sound against SOTA models on T2A, V2A, and VT2A tasks. The \textit{w/ Video-LLaMA caps} row evaluates Omni2Sound's generalization to unseen captions generated by Video-LLaMA \cite{zhang2023video-llama}.}
\label{tab:objective_evaluation_full}

\setlength{\aboverulesep}{1pt}
\setlength{\belowrulesep}{1pt} 

\begin{tabular}{l l cccc cc ccc} 
\toprule
\multirow{2}{*}{\textbf{Task}} & \multirow{2}{*}{\textbf{Method}} & \multicolumn{4}{c}{\textbf{Distribution Matching}} & \multicolumn{2}{c}{\textbf{Audio Quality}} & \multicolumn{3}{c}{\textbf{Modality Alignment}} \\
\cmidrule(lr){3-6} \cmidrule(lr){7-8} \cmidrule(lr){9-11}
& & $\mathrm{\textbf{KL}}$↓ & $\mathrm{\textbf{FD}}$↓ & $\mathrm{\textbf{FAD}}$↓ & $\mathrm{\textbf{FD}_{\textbf{PaSST}}}$↓ & $\mathrm{\textbf{PQ}}$↑ & $\mathrm{\textbf{IS}}$↑ & $\mathrm{\textbf{DS}}$↓ & $\mathrm{\textbf{IB}}$↑ & $\mathrm{\textbf{MS-CLAP}}$↑ \\
\midrule
\multirow{3}{*}{\textbf{T2A}}
& AudioX \cite{tian2025audiox} & \underline{1.68} & 9.04 & \underline{1.42} & 109.94 & \underline{6.37} & \underline{15.15} & - & - & \underline{0.49} \\
& MMAudio \cite{cheng2025mmaudio} & 1.92 & \underline{8.62} & 1.63 & \underline{101.66} & 5.84 & 14.30 & - & - & 0.50 \\
& \cellcolor{cyan!15}\textbf{Omni2Sound (ours)} & \cellcolor{cyan!15}\textbf{1.53} & \cellcolor{cyan!15}\textbf{4.61} & \cellcolor{cyan!15}\textbf{1.01} & \cellcolor{cyan!15}\textbf{60.38} & \cellcolor{cyan!15}\textbf{6.52} & \cellcolor{cyan!15}\textbf{16.41} & \cellcolor{cyan!15}- & \cellcolor{cyan!15}- & \cellcolor{cyan!15}\textbf{0.53} \\
& \textit{\quad w/ Video-LLaMA caps} & 1.60 & 6.92 & 1.23 & 83.91 & 6.38 & 16.01 & - & - & 0.51 \\
\midrule
\multirow{5}{*}{\textbf{V2A}}
& V-AURA \cite{viertola2024vaura} & 2.28 & 16.43 & 2.34 & 245.25 & 5.74 & 10.82 & 0.69 & \underline{0.28} & 0.32 \\
& Frieren \cite{wang2024frieren} & 2.73 & 12.13 & 1.23 & 123.75 & 5.82 & 11.32 & 0.86 & 0.21 & 0.31 \\
& AudioX \cite{tian2025audiox}& 2.96 & 12.73 & 1.42 & 121.82 & \textbf{6.17} & \underline{13.34} & 1.22 & 0.24 & 0.34 \\
& MMAudio \cite{cheng2025mmaudio}& \underline{2.11} & \underline{5.65} & \underline{0.81} & \underline{69.33} & 5.72 & 11.85 & \underline{0.48} & \underline{0.28} & \underline{0.43} \\
& \cellcolor{cyan!15}\textbf{Omni2Sound (ours)} & \cellcolor{cyan!15}\textbf{2.04} & \cellcolor{cyan!15}\textbf{3.41} & \cellcolor{cyan!15}\textbf{0.51} & \cellcolor{cyan!15}\textbf{50.19} & \cellcolor{cyan!15}\underline{6.15} & \cellcolor{cyan!15}\textbf{16.18} & \cellcolor{cyan!15}\textbf{0.47} & \cellcolor{cyan!15}\textbf{0.35} & \cellcolor{cyan!15}\textbf{0.44} \\
\midrule
\multirow{6}{*}{\textbf{VT2A}}
& ThinkSound ($w/o$. CoT) \cite{liu2025thinksound}& \underline{1.60} & 7.41 & 1.10 & 116.08 & \textbf{6.21} & 11.73 & \underline{0.53} & 0.26 & 0.43 \\
& HunyuanVideo-Foley \cite{chen2024hunyuanfoley} & 1.74 & 10.02 & 2.36 & 100.53 & 6.18 & 11.58 & 0.57 & \underline{0.32} & 0.45 \\
& AudioX \cite{tian2025audiox}& 1.59 & 8.29 & 1.24 & 103.37 & 6.17 & \underline{14.94} & 1.23 & 0.26 & \underline{0.49} \\
& MMAudio \cite{cheng2025mmaudio}& 1.63 & \underline{5.28} & \underline{0.91} & \underline{68.44} & 5.84 & 13.44 & \textbf{0.49} & 0.29 & \underline{0.49} \\
& \cellcolor{cyan!15}\textbf{Omni2Sound (ours)} & \cellcolor{cyan!15}\textbf{1.35} & \cellcolor{cyan!15}\textbf{2.95} & \cellcolor{cyan!15}\textbf{0.53} & \cellcolor{cyan!15}\textbf{48.20} & \cellcolor{cyan!15}\underline{6.21} & \cellcolor{cyan!15}\textbf{15.79} & \cellcolor{cyan!15}0.49 & \cellcolor{cyan!15}\textbf{0.34} & \cellcolor{cyan!15}\textbf{0.52} \\
& \textit{\quad w/ Video-LLaMA caps} & 1.56 & 3.37 & 0.66 & 53.73 & 6.11 & 15.74 & 0.50 & 0.34 & 0.49 \\
\bottomrule
\end{tabular}
\end{table*}


\section{\textbf{VGGSound-Omni: Unified Evaluation}}

\label{sec:benchmark}

A key challenge in evaluating unified Video-Text-to-Audio (VT2A) models is the absence of a comprehensive benchmark. The VGGSound test set \cite{chen2020vggsound} provides only sparse event labels and lacks detailed captions. Although VGGSounder \cite{zverev2025vggsounder} improved upon this by correcting and introducing modality labels (e.g., A, V, AV) for fidelity evaluation, it still does not provide human-expert-level captions. To bridge this gap, we construct VGGSound-Omni, a multi-track benchmark derived from the VGGSound test set for evaluating both standard unified and specialized off-screen VT2A generation.

\paragraph{VGGSound-Omni Construction.}
\label{sec:benchmark_gt}
We first establish a high-fidelity, human-level caption set covering all 14,000+ videos as the primary evaluation track. Initial captions are generated using our agentic pipeline (Section~\ref{sec:soundatlas_pipeline}) and then systematically validate through an AI-assisted verification workflow: GPT-5 \cite{openai2025gpt5intro, yang2026tooltree, yang2026evotool} served as an auditor, checking whether the captions semantically covered all ``A'' and ``AV'' labels from VGGSounder \cite{zverev2025vggsounder}. Samples flagged with mismatches are routed for targeted human verification. During this manual audit, we find that most flagged discrepancies stemmed from annotation errors in the VGGSounder data itself (e.g., label redundancy and errors caused by visual interference). After correcting these errors, we establish the final, human-verified captions as the definitive ground truth (GT) for all three tasks (VT2A, V2A, and T2A).

Complementing the primary set, we curate a challenging off-screen track (1,000+ items) from two sources: (i) \textit{Natural events}, filtered from VGGSound for low A-V correspondence (via IB-Score \cite{girdhar2023imagebind} and Desync-Score \cite{cheng2025mmaudio}) while excluding background speech; and (ii) \textit{Synthetic music}, formed by mixing aligned background clips from MusicCaps \cite{agostinelli2023musiccaps}. More details are provided in Appendix~\ref{sec:offscreen_track}.



\begin{table*}[t]
\centering
\footnotesize
\setlength{\tabcolsep}{2.6mm}
\caption{Comparison on the Kling-Audio-Eval: Omni2Sound against SOTA models on T2A, V2A, and VT2A tasks.}
\label{tab:objective_evaluation_kling}
\setlength{\aboverulesep}{2pt}
\setlength{\belowrulesep}{2pt}

\begin{tabular}{l l cccc cc ccc} 
\toprule
\multirow{2}{*}{\textbf{Task}} & \multirow{2}{*}{\textbf{Method}} & \multicolumn{4}{c}{\textbf{Distribution Matching}} & \multicolumn{2}{c}{\textbf{Audio Quality}} & \multicolumn{3}{c}{\textbf{Modality Alignment}} \\
\cmidrule(lr){3-6} \cmidrule(lr){7-8} \cmidrule(lr){9-11}
& & $\mathrm{\textbf{KL}}$↓ & $\mathrm{\textbf{FD}}$↓ & $\mathrm{\textbf{FAD}}$↓ & $\mathrm{\textbf{FD}_{\textbf{PaSST}}}$↓ & $\mathrm{\textbf{PQ}}$↑ & $\mathrm{\textbf{IS}}$↑ & $\mathrm{\textbf{DS}}$↓ & $\mathrm{\textbf{IB}}$↑ & $\mathrm{\textbf{LA-CLAP}}$↑ \\
\midrule

\multirow{3}{*}{\textbf{T2A}}
& AudioX \cite{tian2025audiox}& 2.73 & 19.43 & \underline{3.32} & 171.60 & \underline{5.98} & \textbf{12.15} & - & - & \underline{0.28} \\
& MMAudio \cite{cheng2025mmaudio} & \underline{2.54} & \textbf{11.25} & 5.07 & \textbf{142.71} & 5.54 & 9.28 & - & - & 0.28 \\
& \cellcolor{cyan!15}\textbf{Omni2Sound (ours)} & \cellcolor{cyan!15}\textbf{2.36} & \cellcolor{cyan!15}\underline{11.59} & \cellcolor{cyan!15}\textbf{2.62} & \cellcolor{cyan!15}\underline{147.46} & \cellcolor{cyan!15}\textbf{6.26} & \cellcolor{cyan!15}\underline{11.27} & \cellcolor{cyan!15}- & \cellcolor{cyan!15}- & \cellcolor{cyan!15}\textbf{0.28} \\
\midrule

\multirow{3}{*}{\textbf{V2A}}
& AudioX \cite{tian2025audiox}& 3.13 & 18.90 & 4.01 & 205.48 & \textbf{5.87} & \underline{8.31} & 1.20 & 0.23 & 0.13 \\
& MMAudio \cite{cheng2025mmaudio}& \underline{2.94} & \underline{13.41} & \underline{3.87} & \underline{159.30} & 5.50 & 7.59 & \underline{0.62} & \underline{0.24} & \underline{0.14} \\
& \cellcolor{cyan!15}\textbf{Omni2Sound (ours)} & \cellcolor{cyan!15}\textbf{2.47} & \cellcolor{cyan!15}\textbf{8.78} & \cellcolor{cyan!15}\textbf{2.55} & \cellcolor{cyan!15}\textbf{112.21} & \cellcolor{cyan!15}\underline{5.78} & \cellcolor{cyan!15}\textbf{8.56} & \cellcolor{cyan!15}\textbf{0.57} & \cellcolor{cyan!15}\textbf{0.34} & \cellcolor{cyan!15}\textbf{0.18} \\
\midrule
\multirow{5}{*}{\textbf{VT2A}}
& ThinkSound ($w/o$. CoT) \cite{liu2025thinksound}& 2.53 & 11.99 & 3.52 & 206.93 & 5.77 & 6.09 & 0.66 & 0.22 & 0.19 \\
& HunyuanVideo-Foley \cite{chen2024hunyuanfoley} & \underline{2.13} & \underline{8.06} & 3.58 & \textbf{94.64} & \textbf{6.04} & 8.17 & \textbf{0.55} & \textbf{0.34} & 0.23 \\
& AudioX \cite{tian2025audiox}& 2.39 & 14.26 & \underline{3.16} & 149.37 & 5.97 & \textbf{10.23} & 1.21 & 0.23 & \underline{0.26} \\
& MMAudio \cite{cheng2025mmaudio}& 2.41 & 10.12 & 4.90 & 129.21 & 5.53 & 7.46 & 0.59 & 0.25 & 0.20 \\
& \cellcolor{cyan!15}\textbf{Omni2Sound (ours)} & \cellcolor{cyan!15}\textbf{2.10} & \cellcolor{cyan!15}\textbf{7.60} & \cellcolor{cyan!15}\textbf{2.37} & \cellcolor{cyan!15}\underline{106.55} & \cellcolor{cyan!15}\underline{5.98} & \cellcolor{cyan!15}\underline{8.22} & \cellcolor{cyan!15}\underline{0.58} & \cellcolor{cyan!15}\underline{0.32} & \cellcolor{cyan!15}\textbf{0.26} \\
\bottomrule
\end{tabular}
\vspace{-10pt}
\end{table*}

\section{Experiments}

\subsection{Experiment Settings}
\paragraph{Datasets.}
For T2A backbone pre-training, we use a large-scale corpus comprising the train set of audio datasets such as AudioCaps \cite{kim2019audiocaps}, WavCaps \cite{mei2023wavcaps}, Clotho \cite{drossos2020clotho}, AudioSet \cite{gemmeke2017audio}, VGGSound \cite{chen2020vggsound}, FSD50k \cite{fonseca2022fsd50k}, as well as music datasets including MSD \cite{bertin2011million} and FMA \cite{defferrard2016fma}. To maintain consistency, all audio is segmented into 10-second clips and resampled at 16 kHz. Following this, the model is trained for unified VT2A tasks using our proposed SoundAtlas (Section~\ref{sec:soundatlas_pipeline}) and a high-quality, PQ-score-filtered T-A subset derived from the aforementioned pre-training corpus. More details are provided in Appendix Section \ref{sec:implementation_details}. For evaluation, we compare Omni2Sound with SOTA models on three benchmarks: our proposed VGGSound-Omni (Section~\ref{sec:benchmark}), Kling-Audio-Eval \cite{wang2025audiogenomni}, AudioCaps test set \cite{kim2019audiocaps} and AudioAtlas \cite{wang2025audioatlas}. We ensure that these evaluation benchmarks are strictly disjoint from all data used in our training stages to prevent potential data leakage.

\label{sec:eval_metrics}
\paragraph{Evaluation Metrics.}
\vspace{-10pt}
We implement our objective evaluation using the standardized AV-benchmark toolkit \cite{cheng2025mmaudio} on 8-second clips, following previous work \cite{cheng2025mmaudio}. We assess quality across four critical dimensions \cite{liu2023audioldm}. For \textbf{Distribution Matching}, we measure feature similarity between generated and ground-truth audio using Fréchet Distance (FAD \cite{hershey2017cnn}, $\mathrm{FD}_{\mathrm{PaSST}}$ \cite{koutini2022passt}, FD \cite{kong2020panns}) and Kullback-Leibler divergence (KL, $\mathrm{KL}_{\mathrm{PaSST}}$). \textbf{Audio Quality} is assessed via Inception Scores (IS \cite{salimans2016improved}, $\mathrm{IS}_{\mathrm{PaSST}}$) and Production Quality (PQ \cite{tjandra2025meta}) for aesthetics. \textbf{Semantic Alignment} evaluates text-audio consistency (CLAP \cite{elizalde2023clap}, MS-CLAP \cite{wu2023large}) and video-audio alignment (IB \cite{girdhar2023imagebind}). Finally, \textbf{Temporal Alignment} is measured using the Desynchronization Score (DS) predicted by Synchformer \cite{iashin2024synchformer}. Detailed metric definitions and calculations are provided in the Appendix.

\subsection{Main Results}

\paragraph{Evaluation on VGGSound-Omni.}
We present our main results on VGGSound-Omni benchmark in Table~\ref{tab:objective_evaluation_full}. To ensure a fair comparison, all baseline models are re-evaluated using their official checkpoints and the standardized AV-benchmark \cite{cheng2025mmaudio}, using the same video and text conditions. The results demonstrate that Omni2Sound achieves state-of-the-art performance across all three unified tasks (T2A, V2A, and VT2A) compared to both previous unified VT2A models (AudioX \cite{tian2025audiox}, MMAudio \cite{cheng2025mmaudio}) and specialized models (e.g. ThinkSound \cite{liu2025thinksound}, HunyuanVideo-Foley \cite{chen2024hunyuanfoley}). To further validate Omni2Sound’s generalization beyond our SoundAtlas captioning style, we evaluate it on the same VGGSound test clips but use the Video-LLaMA \cite{zhang2023video-llama} captions from ThinkSound \cite{liu2025thinksound}. As shown in Table~\ref{tab:objective_evaluation_full} (w/ Video-LLaMA caps), while performance slightly degrades, our model’s scores still surpass all baselines, confirming its robustness to unseen captioning styles.

\paragraph{Generalization on Third-Party Benchmarks.}
\vspace{-10pt}

To validate generalization, we evaluate on Kling-Audio-Eval \cite{wang2025audiogenomni} and AudioCaps \cite{kim2019audiocaps} results in Table \ref{tab:objective_evaluation_kling} and Appendix Table \ref{tab:t2a_sota_comparison}. On Kling-Audio-Eval, Omni2Sound remains highly competitive despite the domain gap (YouTube-sourced SoundAtlas vs. Kling's professional video). While trailing HunyuanVideo-Foley \cite{chen2024hunyuanfoley} in some metrics, which is expected given its massive data advantage (100k vs 2k hours), our model consistently outperforms other unified and specialized baselines across all tasks. Furthermore, on AudioCaps, Omni2Sound achieves top-tier performance against specialized T2A models, securing the best scores in distribution metrics ($\mathrm{KL}$, $\mathrm{FD}$) and semantic alignment ($\mathrm{CLAP}=0.36$), while remaining highly competitive in audio quality (PQ) and the $\mathrm{FAD}$ metric.

\paragraph{Subjective Evaluation.}
\label{par:user_study}
To validate perceptual performance, we conduct a human evaluation (detailed in Appendix~\ref{sec:UserStudy}) across three dimensions: Acoustic Fidelity (MOS-Q), Semantic Consistency (MOS-S), and Temporal Synchronization (MOS-T). As shown in Appendix Fig.~\ref{fig:vt2a_scores_comparison}, Omni2Sound outperforms all baselines on both VT2A and V2A tasks. Crucially, these subjective results are highly consistent with the objective metrics in Table~\ref{tab:objective_evaluation_full}, confirming our model's superiority in both generation quality and cross-modal alignment.


\begin{table*}[t]
\centering
\footnotesize
\setlength{\tabcolsep}{2.4mm}
\caption{Ablation study on the Stage 2 multi-task training strategy. TA*/VTA* denotes data from our high-alignment SoundAtlas dataset, while TA/VTA denotes data from a baseline with audio-only captions generated by Gemini 2.5.}
\label{tab:stage2_ablations}
\begin{tabular}{l l cc cccc cccc}
\toprule
\multirow{2}{*}{\textbf{Training Strategy}} &
\multirow{2}{*}{\textbf{$\pi_{T2A}$ : $\pi_{V2A}$ : $\pi_{VTA}$}} &
\multicolumn{2}{c}{\textbf{T2A Task}} & \multicolumn{4}{c}{\textbf{V2A Task}} & \multicolumn{4}{c}{\textbf{VT2A Task}} \\
\cmidrule(lr){3-4} \cmidrule(lr){5-8} \cmidrule(lr){9-12}
& & $\mathrm{\textbf{FAD}}$↓ & $\mathrm{\textbf{FD}}$↓ & $\mathrm{\textbf{FAD}}$↓ & $\mathrm{\textbf{FD}}$↓ & $\mathrm{\textbf{DS}}$↓ & $\mathrm{\textbf{IB}}$↑ & $\mathrm{\textbf{FAD}}$↓ & $\mathrm{\textbf{FD}}$↓ & $\mathrm{\textbf{DS}}$↓ & $\mathrm{\textbf{IB}}$↑ \\
\midrule

TA+VA & 0.20 : 0.80 : 0.00 & 1.36 & 5.52 & 0.56 & 4.13 & 0.50 & 0.33 & - & - & - & - \\
TA+VA & 0.40 : 0.60 : 0.00 & 1.06 & 4.62 & 0.62 & 4.63 & 0.52 & 0.32 & - & - & - & - \\

TA*+VA+VTA* & 0.10 : 0.35 : 0.55 & 0.94 & 4.22 & 0.57 & 3.61 & 0.49 & 0.33 & 0.53 & 2.83 & 0.51 & 0.32 \\
TA+VA+VTA & 0.20 : 0.30 : 0.50 & 1.13 & 4.68 & 0.56 & 4.22 & 0.50 & 0.32 & 0.62 & 3.51 & 0.51 & 0.33 \\
\bottomrule
\end{tabular}
\end{table*}

\begin{table}[t]
\centering
\footnotesize
\setlength{\tabcolsep}{2.5mm} 

\caption{Ablation study on our progressive multi-task training. We compare our full S1 $\rightarrow$ S2 $\rightarrow$ S3 model against three baselines (S2, S1 $\rightarrow$ S2, and S1 $\rightarrow$ [S2+S3]). All models are trained for the same total 1.2M steps.}
\label{tab:ablation_stages}

\begin{tabular}{l c c c c c} 
\toprule
\textbf{Task} & \textbf{Method} & $\mathrm{\textbf{FAD}}$↓ & $\mathrm{\textbf{FD}}$↓ & $ \mathrm{\textbf{DS}}$↓ & $\mathrm{\textbf{IB}}$↑ \\
\midrule
\multirow{4}{*}{\textbf{T2A}}
& S2 & 1.22 & 5.88 & - & - \\
& S1 $\rightarrow$ S2 & 0.94 & 4.62 & - & - \\
& S1 $\rightarrow$ [S2+S3] & 1.11 & 4.45 & - & - \\
& \textbf{S1 $\rightarrow$ S2 $\rightarrow$ S3} & 1.01 & 4.61 & - & - \\ 
\midrule
\multirow{4}{*}{\textbf{V2A}}
& S2 & 0.68 & 4.70 & 0.47 & 0.33 \\
& S1 $\rightarrow$ S2 & 0.57 & 3.61 & 0.49 & 0.33 \\
& S1 $\rightarrow$ [S2+S3] & 0.60 & 3.81 & 0.47 & 0.34 \\
& \textbf{S1 $\rightarrow$ S2 $\rightarrow$ S3} & 0.51 & 3.41 & 0.47 & 0.35 \\ 
\midrule
\multirow{4}{*}{\textbf{VT2A}}
& S2 & 0.63 & 4.40 & 0.49 & 0.33 \\
& S1 $\rightarrow$ S2 & 0.53 & 2.83 & 0.51 & 0.32 \\
& S1 $\rightarrow$ [S2+S3] & 0.61 & 3.27 & 0.50 & 0.33 \\
& \textbf{S1 $\rightarrow$ S2 $\rightarrow$ S3} & 0.53 & 2.95 & 0.49 & 0.34 \\ 
\bottomrule
\end{tabular}
\end{table}

\subsection{Ablation Studies}
\label{sec:experiments} 

We first analyze the multi-task training dynamics in Table~\ref{tab:stage2_ablations} to demonstrate how high-quality data resolves task competition, and then use Table~\ref{tab:ablation_stages} to prove the necessity of our three-stage progressive training schedule.

\paragraph{High-Quality VT2A Data as a Critical Bridge.}
We first investigate the Cross-Task Competition between V2A and T2A, which persists even when models are based on the T2A pretraining from Stage 1. As shown in Table~\ref{tab:stage2_ablations} (rows 1-2), a naive joint training of V2A and T2A results in a severe trade-off. Increasing the T2A sampling ratio ($\pi_{T2A}$) from 0.20 to 0.40 improves T2A performance (FAD 1.36 $\rightarrow$ 1.06) but simultaneously degrades V2A generation (FAD 0.56 $\rightarrow$ 0.62), preventing simultaneous optimization. Our insight is that this conflict is resolved by introducing high-quality VT2A data as a critical bridge. This hypothesis is validated in row 3, which introduces our SoundAtlas data (denoted by TA* and VTA*). The results show a dramatic performance boost, achieving the best metrics across all tasks (e.g., T2A FAD 0.94, V2A FD 3.61, VT2A FD 2.83). This confirms that the high V-A-T alignment in SoundAtlas is essential for resolving the V2A-T2A competition and fostering a cooperative dynamic.

To further emphasize that this bridging effect is contingent on data quality, we provide a comparison in row 4. Here, we use standard-quality data (TA/VTA), where captions were generated by Gemini-2.5 using only the audio modality. Although the VT2A task is present, the poor V-T-A alignment fails to resolve the competition, and performance is still severely compromised (e.g., T2A FAD 1.13), far underperforming the SoundAtlas-driven model. This comparison proves that it is not merely the VT2A task, but the high-fidelity alignment of the bridge data, that is essential. This high quality enables data efficiency: the T2A ratio can be dropped to $\pi_{T2A}=0.1$ while achieving SOTA T2A performance, mitigating resource contention as designed.

\paragraph{Necessity of the Progressive Three-Stage Schedule.}
Next, we demonstrate the necessity of our full progressive schedule in Table~\ref{tab:ablation_stages}. We compare our full S1 $\rightarrow$ S2 $\rightarrow$ S3 pipeline against three baselines, all trained for the same total steps on SoundAtlas data. First, comparing the S2 only model with the S1 $\rightarrow$ S2 model confirms the value of the Stage 1 generative prior. Without S1, the S2 only model fails to converge well, showing poor quality (T2A FAD 1.22, V2A FAD 0.68). The S1 $\rightarrow$ S2 model, benefiting from the pretraining, significantly boosts generation quality (T2A FAD 0.94, V2A FAD 0.57) and resolves the Cross-Task Competition. However, this model still suffers from Intra-Task Competition (modality bias), as evidenced by its weaker A-V synchronization (V2A DS 0.49). Second, we validate our crucial hypothesis that Stage 3 must be decoupled. The S1 $\rightarrow$ [S2+S3] baseline, which merges the S3 robustness augmentations directly into S2, destabilizes the fragile optimization process. While it maintains A-V synchronization (V2A DS 0.47), introducing these augmentations prematurely harms the generative quality achieved in S2, leading to a clear degradation in FAD/FD scores (e.g., V2A FAD 0.60, VT2A FAD 0.61). 

Finally, our full S1 $\rightarrow$ S2 $\rightarrow$ S3 model resolves both challenges. As established in our method, S3 has two complementary goals: mitigating the text bias (via Text Dropout) and the video bias (via Off-screen Synthesis). The main results in Table~\ref{tab:ablation_stages} confirm the first goal: the full S3 model enhances cross-modal consistency (V2A DS 0.49 $\rightarrow$ 0.47) while achieving the highest overall generation quality (V2A FAD 0.51). To validate the second goal—improving faithfulness against a video bias—we conduct a targeted evaluation on our VGGSound-Omni off-screen track, presented in Table~\ref{tab:ablation_offscreen_track}. This table compares the S1$\rightarrow$S2 baseline against our full model, showing the S3 augmentations yield superior audio quality and improved objective text-audio alignment. This gain in faithfulness is further confirmed by a subjective preference test using an MLLM-as-Judge (evaluating text-audio faithfulness on a 1-to-5 scale).

\begin{table}[t]
\centering
\footnotesize
\caption{VT2A evaluation on VGGSound-Omni off-screen track.
We compare S1$\rightarrow$S2 against our full S1$\rightarrow$S2$\rightarrow$S3 model to validate \textit{Off-screen Synthesis} augmentation.}
\label{tab:ablation_offscreen_track} 
\setlength{\tabcolsep}{2.2mm} 
\begin{tabular}{l c c c c}
\toprule
\textbf{Method} & \textbf{FAD}↓ & \textbf{KL}↓ & \textbf{LA-CLAP}↑ & \textbf{Win Rate}↑ \\
\midrule
S1 $\rightarrow$ S2 & 0.97 & 1.46 & 0.31 & 46.8\% \\
\textbf{S1 $\rightarrow$ S2 $\rightarrow$ S3} & 0.85 & 1.39 & 0.32 & 53.2\% \\
\bottomrule
\end{tabular}
\end{table}

\section{Conclusion}
In this work, we addressed the foundational challenges of unified video-text-to-audio (VT2A) generation: data scarcity and cross-task competition. We introduce a three-part contribution: SoundAtlas, the first large-scale, human-expert-level audio caption dataset; Omni2Sound, a unified model featuring a three-stage progressive schedule to resolve task competition; and VGGSound-Omni, a comprehensive benchmark for unified VT2A evaluation. Our experiments demonstrate that this approach effectively resolves cross-task and intra-task competition and enables Omni2Sound to achieve unified state-of-the-art performance.

\section*{Acknowledgments}
This work is supported by the Fundamental and Interdisciplinary Disciplines Breakthrough Plan of the Ministry of Education of China (JYB2025XDXM101), the National Natural Science Foundation of China (62550004, U24A20342, U25B6003, 92570001), and the Australian Research Council (DP260100218).

\bigskip
{
   \small
    \bibliographystyle{unsrt} 
    \bibliography{main}
}

\appendix 

\clearpage
\setcounter{page}{1}
\maketitlesupplementary


\paragraph{Overview} This document provides technical details, evaluation protocols, and extended experimental analyses. We begin with the \textbf{Cost Analysis} in Section \ref{sec:cost_analysis}, validating SoundAtlas as a scalable and cost-effective pipeline. We then provide the exact \textbf{Audio Caption Prompt Instructions} in Section \ref{sec:prompt}, followed by detailed \textbf{Evaluation Protocols} to compare the quality of Audio Caption Datasets in Section \ref{sec:llm_judge_validation} and the detailed construction of the \textbf{Off-Screen Benchmark Track} in Section \ref{sec:offscreen_track}. Furthermore, we demonstrate the model's \textbf{Generalization Capabilities on third-party benchmarks} in Section \ref{sec:Generalization} and elaborate on the \textbf{User Study} in Section \ref{sec:UserStudy}. Section \ref{sec:implementation_details} outlines the \textbf{Implementation Details}, including model configurations and training data composition. Section \ref{sec:eval_metrics} defines the \textbf{Objective Evaluation Metrics on Generation Audio} used throughout the paper. \textbf{Qualitative results} can be found in static HTML file.

\section{Cost Analysis on Audio Captioning}
\label{sec:cost_analysis}
While Gemini 2.5 Pro \cite{team2023gemini} represents a milestone as a native multimodal foundation model, utilizing it directly for large-scale video-grounded audio captioning proves economically unsustainable. As quantified in Table \ref{tab:cost_analysis}, using Gemini's standard API pricing, a naive implementation—processing raw video frames alongside audio ($V+A$)—incurs a prohibitive expenditure of \$10,275 USD per 1M samples. This figure is derived from the token consumption of a 10-second sample: the input aggregates to 3,820 tokens (comprising 1,000 instruction, 320 audio, and 2,500 visual tokens), while the full chain-of-thought generation requires $\sim$550 output tokens. Crucially, this naive approach suffers from an inherent visual bias, as shown in Figure \ref{fig:small} in main paper.

To address these challenges, our SoundAtlas pipeline employs three strategic optimizations. First, we implement \textit{Vision-to-Language Compression}. This strategy replaces expensive raw video with a concise video caption $c_v$, eliminating the large $\sim$2,500 token visual overhead (Table \ref{tab:cost_analysis}, Row 2) and effectively mitigating the visual modality bias. Second, we enforce \textit{Restricted Reasoning}, capping the generation output at $\sim$160 tokens (Table \ref{tab:cost_analysis}, Row 3). Finally, we utilize a \textit{Junior-Senior Agent Handoff} that defaults to the cost-effective Flash model $G_{\text{junior}}$ for the majority of samples, reserving the Senior agent ($G_{\text{senior}}$) solely for complex cases. As shown in Table \ref{tab:cost_analysis}, while the standalone Flash model offers the lowest theoretical cost (\$1,026), our hybrid pipeline strikes a balance between quality and efficiency, reducing the initial expenditure of \$10,275 to approximately \$2,000 per million samples.

\begin{table*}[t]
\centering
\setcounter{table}{5}
\footnotesize
\caption{Cost Analysis on Audio Captioning with Gemini 2.5. We compare the inference costs for processing one million 10-second samples. The table demonstrates a step-by-step ablation path: removing raw video (Row 2), restricting reasoning with vision-to-language compression (Row 3), and switching to the Flash model (Row 4) progressively reduces costs from \$10,275 to \$1,026.}
\label{tab:cost_analysis}
\setlength{\tabcolsep}{2.5mm} 
\begin{tabular}{lccccr}
\toprule
\textbf{Model Configuration} & \textbf{Input Modality} & \textbf{Input Token Num.} & \textbf{Output Token Num.} & \textbf{Est. Cost (USD / 1M Samples)} \\
\midrule
Gemini 2.5 Pro (Thinking-Full) & T + V + A & 3,820 & 550 & \$10,275.00 \\
Gemini 2.5 Pro (Thinking-Full) & T + A & 1,340 & 550 & \$7,175.00 \\
Gemini 2.5 Pro (Thinking-128) & T + A & 1,340 & 160 & \$3,275.00 \\
Gemini 2.5 Flash (Thinking-128) & T + A & 1,340 & 160 & \$1,026.00 \\
\bottomrule
\end{tabular}
   \vspace{-10pt}
\end{table*}


\begin{table}[t]
\centering
\footnotesize
\setlength{\tabcolsep}{1.2mm} 
\caption{Comparison of the generation performance on unified VT2A models and T2A models on Audiocaps test set.}
\label{tab:t2a_sota_comparison}
\begin{tabular}{l ccc cc} 
\toprule
\textbf{Method} & $\mathrm{\textbf{KL}}$↓ & $\mathrm{\textbf{FD}}$↓ & $\mathrm{\textbf{FAD}}$↓ & $\mathrm{\textbf{PQ}}$↑ & $\mathrm{\textbf{LA-CLAP}}$↑ \\
\midrule
AudioLDM 2-L \cite{liu2024audioldm2} & 1.73 & 34.21 & 2.26 & \textbf{5.93} & 0.24 \\
TANGO 2 \cite{majumder2024tango2} & \textbf{1.19} & 15.92 & 3.17 & 5.82 & \underline{0.35} \\
Make-An-Audio 2 \cite{huang2023makeanaudio2_placeholder} & 1.38 & 15.34 & \underline{1.46} & 5.64 & 0.25 \\
GenAU-Large \cite{haji-ali2024genau} & 1.42 & 16.92 & \textbf{1.32} & 5.52 & 0.26 \\
\midrule
MMAudio \cite{cheng2025mmaudio} & 1.43 & \underline{13.78} & 2.92 & 5.30 & 0.29 \\
AudioX \cite{tian2025audiox} & 1.55 & 17.10 & 2.65 & 5.81 & 0.31 \\
\cellcolor{cyan!15}\textbf{Omni2Sound (Ours)} & \cellcolor{cyan!15}\underline{1.35} & \cellcolor{cyan!15}\textbf{11.42} & \cellcolor{cyan!15}1.74 & \cellcolor{cyan!15}\underline{5.84} & \cellcolor{cyan!15}\textbf{0.36} \\
\bottomrule
\end{tabular}
\end{table}

\section{Audio Caption Prompt Instructions}
\label{sec:prompt}
As illustrated in Figure \ref{fig:audio_prompt}, we present the audio captioning system prompt employed in our agentic annotation pipeline to construct the \textit{SoundAtlas} dataset. 

\section{Audio Caption Dataset Comparison}
\label{sec:llm_judge_validation}
We provide the detailed scoring process for both MLLM-as-a-judge and Human Expert Evaluation on different audio caption datasets in Table \ref{tab:dataset_clap} and \ref{human_vs_machine_5col_new} of main paper. The evaluation methodology consists of two stages: (1) absolute scoring based on the specific linguistic criteria defined below, and (2) a comparative win-rate calculation derived from these scores.

\paragraph{Subjective Evaluation Protocol.}
We formulate a standardized scoring protocol for both MLLM and human evaluators, focusing on two distinct dimensions of modality alignment.

\noindent\textbf{1. Semantic Alignment (MOS-S, Scale 1-4).} This metric assesses both \textit{Accuracy} (factuality of sound events) and \textit{Detail} (precision of adjectives). The scale is defined as: (1) Factually incorrect/Brief; (2) Mostly incorrect/Brief; (3) Minor errors/Detailed (but visually redundant); and (4) Error-free and Detailed (strictly audio-centric).

\noindent\textbf{2. Temporal Alignment (MOS-T, Scale 1-3).} This evaluates whether the chronological order of described events matches the audio stream. The scale ranges from (1) Disordered, (2) Partially Correct, to (3) Perfectly Ordered. Samples with constant or stationary sounds (lacking distinct temporal events) are marked as \texttt{N/A} and excluded from this metric.

\vspace{0.5em} 
\noindent\textbf{Human Evaluation Setup.}
To complement and validate our automated evaluation, we conducted a dedicated human expert evaluation based on the aforementioned protocol. We randomly sampled a subset of 100 instances from the evaluation corpus used in the MLLM-as-a-judge benchmark. We recruited five expert annotators with professional backgrounds in audio-visual analysis to assess these samples independently. To ensure robustness and mitigate individual bias, the final score for each item is derived by calculating the average rating across the five evaluators. For reference, the user study interface is illustrated in Figure \ref{fig:cap_evaluation}.

\noindent\textbf{Win Rate Calculation.}
We adopt a general pairwise comparison paradigm. For each evaluation set, a target model is compared against an opposing method. The Mean Win Rate (MWR) for any given model is derived by aggregating the outcomes of all its pairwise comparisons:
\begin{equation}
\text{MWR} = \frac{N_{\text{win}} + 0.5 \times N_{\text{tie}}}{N_{\text{total}}}
\end{equation}
where $N_{\text{win}}$, $N_{\text{tie}}$, and $N_{\text{total}}$ denote the number of wins (scoring 1.0), ties (scoring 0.5), and total pairwise comparisons involving that model, respectively. 

\section{Off-Screen Track of VGGSound-Omni}
\label{sec:offscreen_track}

We introduce a dedicated Off-Screen Audio-Generation Track of VGGSound-Omni. This subset specifically evaluates the model's capacity to handle non-depicted audio sources and is constructed through two distinct pipelines: (i) a \textit{Natural Off-screen Events} subset sourced from the original test set; and (ii) a \textit{Synthetic Music} subset focusing on background music (BGM) generation.

\paragraph{Natural Off-screen Events.}
We construct the \textit{Natural Events} subset by identifying VGGSound clips that inherently contain off-screen audio cues. The curation involves a rigorous three-step filtering pipeline. First, regarding Metadata \& Modality, we ensure acoustic purity by excluding samples with pre-existing background music, static imagery, or voice-overs. Crucially, we filter out videos containing vision-only (``V'') labels, retaining only those with Audio-Visual (``AV'') or Audio-only (``A'') modalities. Second, for Complexity \& Consistency, we limit scene complexity to a maximum of 6 labels. To capture "natural" off-screen scenarios, we filter based on the AV Ratio—defined as the proportion of ``AV'' labels relative to the total label count. We explicitly select samples where this ratio falls within $[0.25, 0.80]$, ensuring that the audio content is not perfectly aligned with the visual stream (i.e., low A-V correspondence). Finally, we apply Distribution Balancing to mitigate the over-representation of common classes, restricting the proportion of speech to 20\%.

\begin{figure*}[t] 
    \centering
    
    \begin{subfigure}[b]{0.48\linewidth}
        \centering
        \includegraphics[width=\linewidth]{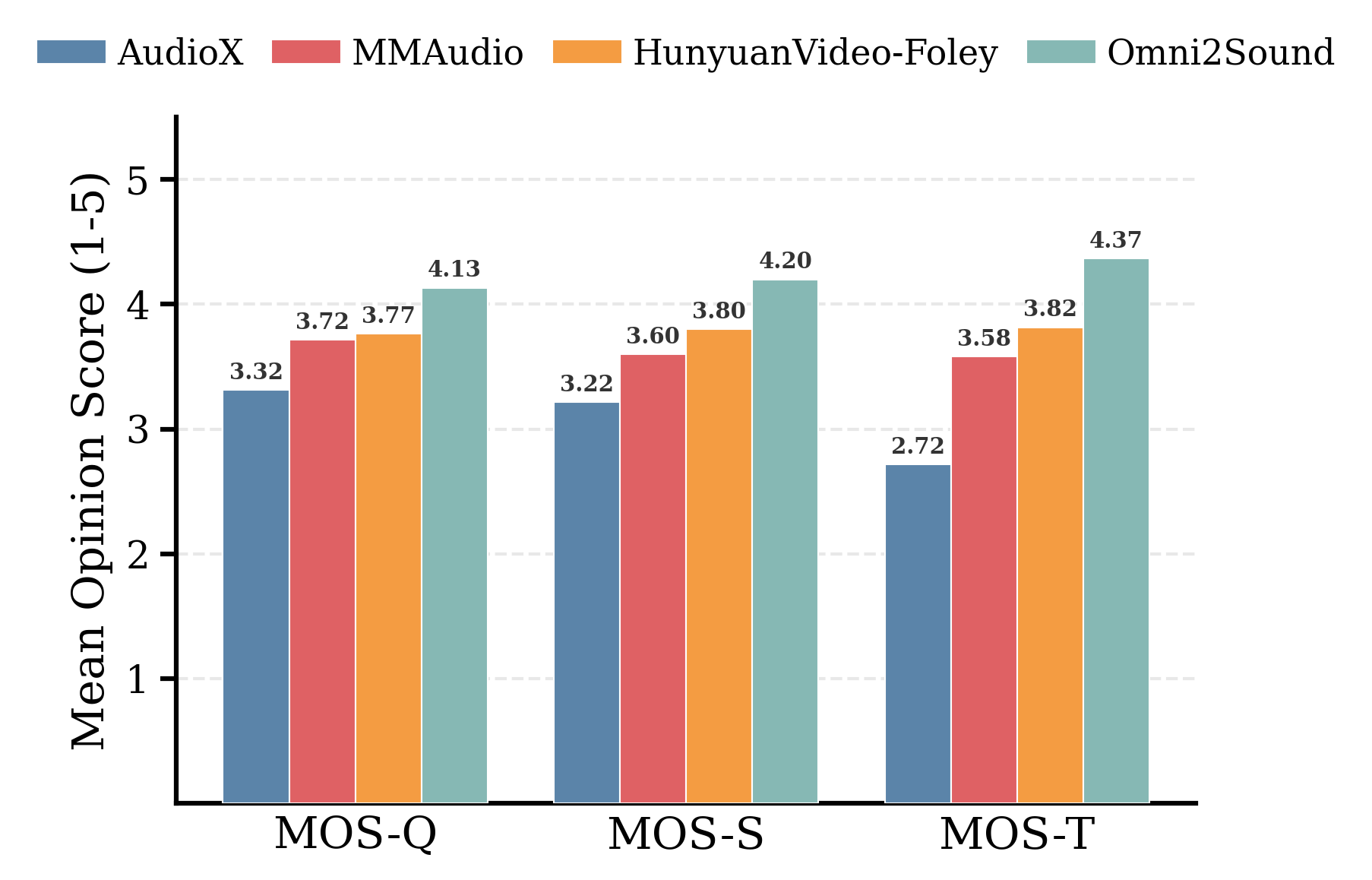}
        \caption{Subjective Evaluation Results on VT2A Task} 
        \label{fig:sub_a}
    \end{subfigure}
    \hfill 
    \begin{subfigure}[b]{0.5\linewidth}
        \centering
        \includegraphics[width=0.9\linewidth]{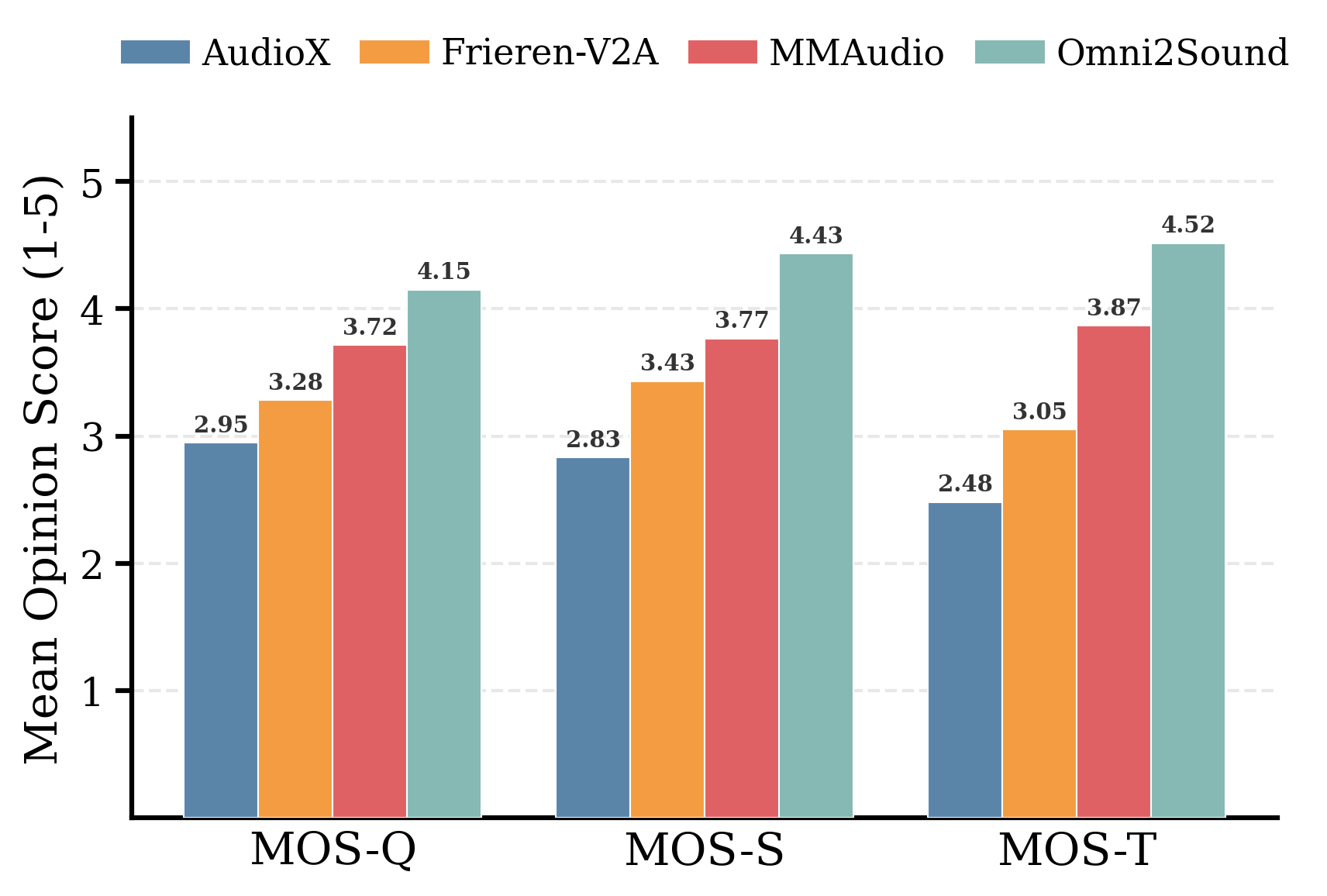}
        \caption{Subjective Evaluation Results on V2A Task} 
        \label{fig:sub_b}
    \end{subfigure}
    
    \caption{Subjective Evaluation Results on VGGSound-Omni. We report Mean Opinion Scores (MOS) on a 1-5 scale across three dimensions: Acoustic Quality (MOS-Q), Semantic Alignment (MOS-S), and Temporal Alignment (MOS-T). Omni2Sound consistently outperforms competitive baselines (AudioX, MMAudio, HunyuanVideo-Foley, Frienren-V2A) across all perceptual metrics on both VT2A and VT2A tasks, validating its superior generation fidelity and alignment.}
    \label{fig:vt2a_scores_comparison}
    \vspace{-15pt}
\end{figure*}

\paragraph{Synthetic Music Augmentation.}
To address the high demand for Background Music (BGM) generation, we create a \textit{Synthetic Music} subset by mixing semantically aligned MusicCaps~\cite{agostinelli2023musiccaps} clips into a pool of high-fidelity videos. This process follows a two-stage procedure. In the Base Selection stage, we first select a "clean" video pool by strictly requiring a 100\% AV label ratio and filtering for high alignment (ImageBind $\ge 0.30$, Desync $< 0.55$), ensuring all original acoustic events are visually manifest. Subsequently, during Semantic Mixing, we augment these videos with background music tracks. To guarantee semantic coherence, we utilize GPT to retrieve the most congruent music track from a random candidate batch of 50 samples based on the video context. Ground-truth captions are updated to reflect this acoustic addition.

\vspace{-12pt}
\paragraph{Comparison with Concurrent Work.}
We acknowledge the pioneering work of VinTAGe-Bench~\cite{bain2025vintage} in synthetic robustness evaluation. However, the off-screen subset of our VGGSound-Omni benchmark extends this direction in three critical dimensions. First, in terms of \textbf{Realism}, by leveraging VGGSounder~\cite{zverev2025vggsounder} metadata, our natural subset is primarily sourced from real-world off-screen audio events rather than relying solely on synthetic mixes. Second, regarding \textbf{Scale}, our benchmark is significantly larger, providing 1,613 evaluation items compared to the 212 basic videos of VinTAGe-Bench. Third, regarding \textbf{Scope}, we include a dedicated \textit{Synthetic Music (BGM)} track, addressing a critical, high-demand scenario often overlooked in standard environmental sound benchmarks.

\section{\texorpdfstring{\fontsize{11}
{13}\selectfont \textbf{Generalization on Third-Party Benchmarks.}}{Generalization on Third-Party Benchmarks.}}
\label{sec:Generalization}
To further validate our model's generalization and mitigate potential biases from our self-constructed benchmark, we evaluate it on the Kling-Audio-Eval \cite{wang2025audiogenomni} and Audiocaps test set \cite{kim2019audiocaps}.  In Table~\ref{tab:objective_evaluation_kling}, on the Kling-Audio-Eval benchmark, Omni2Sound remains highly competitive, despite a significant data scale and distribution gap (our YouTube-sourced SoundAtlas vs. Kling's professional video/Foley). While HunyuanVideo-Foley \cite{chen2024hunyuanfoley} leads on several metrics, this is expected given its massive 100k-hour internal dataset, which is tens of times larger than our SoundAtlas filter derived from VGGSound and AudioSet. Nevertheless, Omni2Sound consistently outperforms all other strong baselines (e.g., MMAudio, AudioX, and ThinkSound) across V2A and VT2A tasks, demonstrating strong generalization as the SOTA or second-best method. In Table \ref{tab:t2a_sota_comparison}, on the Audiocaps test set, we compare Omni2Sound against specialized SOTA T2A models. The results show our unified model achieves top-tier performance, attaining the best scores in key distribution metrics ($\mathrm{KL}$, $\mathrm{FD}$) and semantic alignment ($\mathrm{CLAP}=0.36$), while remaining highly competitive in audio quality (PQ) and the $\mathrm{FAD}$ metric.

\section{User Study}
\label{sec:UserStudy}
We conduct a comprehensive user study on the VGGSound-Omni benchmark to validate Omni2Sound against top baselines (four methods in total). Given the density of comparisons involved, we structure VT2A and V2A as independent evaluation tracks to mitigate evaluator fatigue. We recruit a total of 16 expert evaluators, who are evenly distributed across the two independent tasks. Each participant evaluates 20 random samples (80 comparisons) within their assigned track. Samples from the same source are grouped with randomized method order to maintain blinding. In total, 1280 responses per metric are collected.

\paragraph{Subjective Evaluation Metrics.}
\label{sec:eval_metrics2}
Our final evaluation utilizes a multi-dimensional Mean Opinion Score (MOS) protocol, where expert human evaluators assess the generated audio across three distinct criteria. All scores are normalized to a 5-point Likert scale (1: Poor/Misaligned; 5: Excellent/Perfectly Aligned).

\begin{itemize}
\item \textbf{MOS-Q: Acoustic Fidelity (Quality).} This metric assesses the intrinsic acoustic quality and perceptual realism of the generated sound, independent of the conditioning inputs. Evaluators focus on auditory naturalness, clarity, and the absence of technical artifacts (e.g., distortion, noise, mixing comfort).

\item \textbf{MOS-S: Semantic Consistency (Alignment).} This quantifies the perceptual fidelity between the content of the generated audio and the semantic information conveyed by the conditioning modalities (video frames and textual captions). Evaluation centers on whether the generated sound event's category and characteristics logically correspond to the depicted visual and textual context.

\item \textbf{MOS-T: Temporal Synchronization (Alignment).} This assesses the temporal accuracy of the acoustic events against the visual stream. Evaluators specifically check the precision of sound onset, offset, and duration, ensuring tight synchronization with the corresponding visual event timing.
\end{itemize}

\begin{figure*}[h] 
    \centering
    
    \begin{minipage}{0.48\textwidth} 
        \centering
        \includegraphics[width=\linewidth]{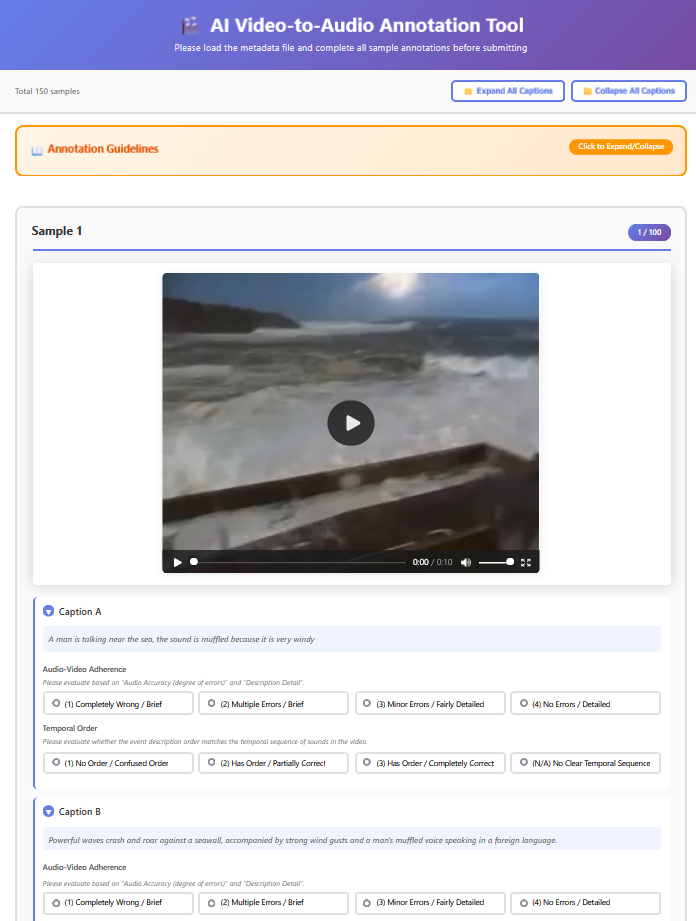}
        \caption{User study interface for human evaluation across different audio generation models.}
        \label{fig:cap_evaluation}
    \end{minipage}
    \hfill 
    \begin{minipage}{0.48\textwidth}
        \centering
        \includegraphics[width=\linewidth]{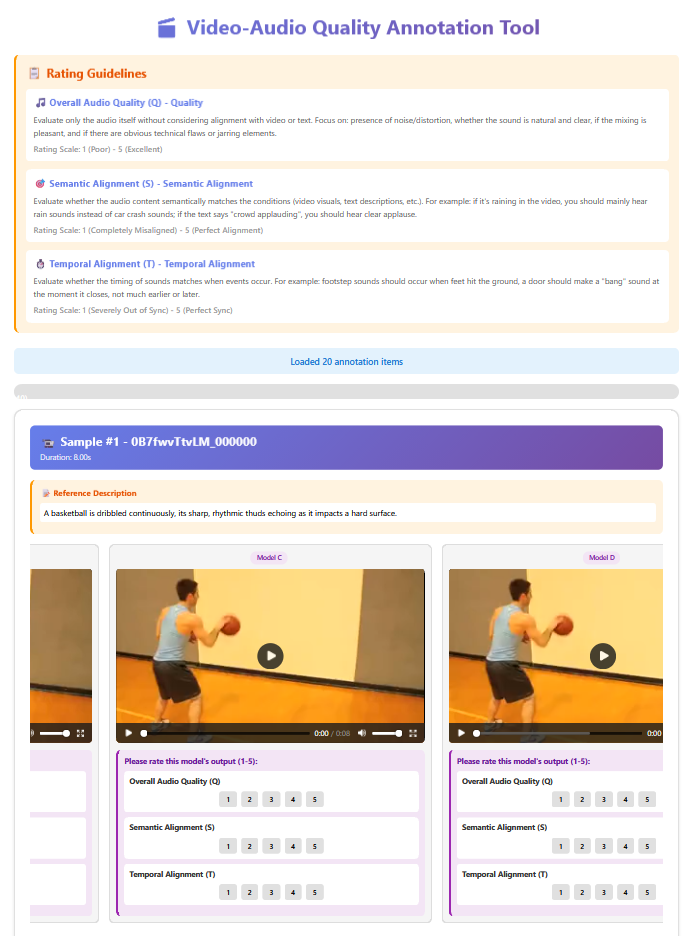}
        \caption{User study interface for human evaluation across different automatic audio captioning datasets.}
        \label{fig:audio_evaluation}
    \end{minipage}
    
\end{figure*}

The results, summarized in Figure \ref{fig:vt2a_scores_comparison}, demonstrate that Omni2Sound outperforms all baselines across the three subjective metrics: MOS-Q, MOS-S, and MOS-T on both VT2A and V2A tasks. This strong alignment between human preference in Figure \ref{fig:vt2a_scores_comparison} and the objective metrics presented in Table \ref{tab:objective_evaluation_full} in main paper validates the effectiveness of our proposed data construction and training pipeline. For reference, the user study interface is illustrated in Figure \ref{fig:audio_evaluation}.

\section{Implementation Details}
\label{sec:implementation_details}

\paragraph{Model Configuration.}
Following Stable Audio~\cite{evans2024stableaudioopen}, our diffusion model adopts \textbf{a} Diffusion Transformer (DiT) architecture within a Latent Diffusion Model (LDM) paradigm. The diffusion backbone consists of a DiT with 24 layers, 24 attention heads, and a hidden dimension of 1536. We employ cross-attention mechanisms to inject semantic conditions (e.g., FLAN-T5 and CLIP embeddings) and Adaptive Layer Normalization (AdaLN) to integrate temporal signals, as detailed in Section \ref{sec:foundation_model}. Both the conditional token dimension and the global condition embedding dimension are 1024. Finally, for audio compression, we train a Variational Autoencoder (VAE) from scratch based on the wav Audio VAE architecture~\cite{evans2024stableaudioopen}, operating at a 16kHz sampling rate. With strides of $[4, 4, 4, 10]$, the encoder achieves a total downsampling ratio of 640, mapping mono waveforms into a compact 64-dimensional latent space. To ensure high-fidelity reconstruction, we utilize Snake activations throughout the network.

\paragraph{Training Data.}
\label{sec:training_data} 
For T2A backbone pre-training, we use a large-scale corpus comprising the train set of audio datasets such as AudioCaps \cite{kim2019audiocaps}, WavCaps \cite{mei2023wavcaps}, Clotho \cite{drossos2020clotho}, AudioSet \cite{gemmeke2017audio}, VGGSound \cite{chen2020vggsound}, FSD50k \cite{fonseca2022fsd50k}, as well as music datasets including MSD \cite{bertin2011million} and FMA \cite{defferrard2016fma}. All audio signals are standardized to a mono-channel format at 16kHz. To accommodate fixed-size diffusion inputs, we normalize clips to a uniform 10-second duration: samples exceeding this length undergo right cropping, while shorter samples are right-padded with silence.

Subsequently, the model is fine-tuned for unified multimodal tasks using our proposed SoundAtlas. Constructed following the pipeline \textbf{detailed in} Section~\ref{sec:benchmark}, this dataset comprises 470k high-quality V-A-T pairs, sourced from 140k VGGSound and 330k AudioSet samples. Notably, the AudioSet subset is strictly curated: starting from the original 2M corpus, we first applied a preliminary filtration to exclude all speech- and music-related categories, resulting in a candidate pool of 450k sound samples. These candidates then underwent our A-V consistency routing and verification pipeline to yield the final 330k high-fidelity pairs. For T2A task fine-tuning, we augment the training with T-A pairs from SoundAtlas as well as a high-fidelity subset of the pre-training corpus, filtered by strict quality thresholds: requiring a CLAP score greater than 0.35 and a PQ score exceeding 6.0.

\section{Objective Evaluation Metrics.}
\label{sec:eval_metrics}
We implement our objective evaluation metrics using the standardized AV-benchmark toolkit \cite{cheng2025mmaudio}. All samples are generated under the same video and text conditions and evaluated in 8-second clips, following previous work \cite{cheng2025mmaudio}. Following common practice \cite{liu2023audioldm}, we assess the quality of the generation in four critical dimensions. For Distribution Matching, we measure the similarity in feature distribution between generated and ground-truth audio. We compute the Fr\'echet Distance using the VGGish (FAD) \cite{hershey2017cnn} and PaSST ($\mathrm{FD}_{\mathrm{PaSST}}$) \cite{koutini2022passt}  embeddings, as well as the Fr\'echet Audio Distance using PANNs (FD) \cite{kong2020panns}. We also report the Kullback-Leibler divergence using PANNs (KL) and PaSST ($\mathrm{KL}_{\mathrm{PaSST}}$) classifiers. For Audio Quality, we assess the quality of the generation using the Inception Score \cite{salimans2016improved}, calculated with both the PANNs (IS) and PaSST ($\mathrm{IS}_{\mathrm{PaSST}}$) classifiers. For Semantic Alignment, we evaluate text-audio consistency using LAION CLAP (CLAP) \cite{elizalde2023clap} and Microsoft CLAP (MS-CLAP) \cite{wu2023large} scores, and video-audio alignment using ImageBind score (IB) \cite{girdhar2023imagebind} as cosine similarity between video and audio embeddings. Finally, for Temporal Alignment, we assess audio-visual synchrony using the DS metric predicted by Synchformer \cite{iashin2024synchformer}.

\begin{figure*}[h!]
\centering
\begin{tcolorbox}[
colback=bg-gray,
colframe=black,
arc=8pt,
boxrule=0.8pt,
width=\textwidth,
boxsep=8pt,
left=10pt, right=10pt, top=10pt, bottom=10pt,
fontupper=\sffamily\small 
]

\begin{center}
\textbf{\Large Audio Captioning Instruction for SoundAtlas}
\end{center}
\vspace{0.5em}

\textbf{\large Roles and Tasks}
\vspace{0.3em}

You are an experienced audio content analyst skilled in describing soundscapes through detailed, multi-dimensional natural language. Given an audio clip (\textit{a}) and its corresponding video descriptions (\textit{$T_v$}), identify and describe all relevant auditory elements in chronological order, then write a rich audio description that faithfully and dynamically reflects the scene.

\vspace{0.8em}

\textbf{\large Annotation Dimensions}
\begin{enumerate}[leftmargin=*, nosep, label=\textbf{\arabic*.}]
\item \textbf{Primary Sound Information}
\begin{itemize}[leftmargin=1em, topsep=2pt, itemsep=0pt]
\item \textbf{Humans/Animals:} speech (talking, shouting), movements (footsteps). \textit{Note: Do not transcribe words/lyrics; describe voice characteristics.}
\item \textbf{Objects:} traffic, office sounds, battlefield, tools.
\item \textbf{Characteristics:} Gender/age, language, quantity (monologue/turn-taking), emotional tone, voice qualities.
\end{itemize}

\item \textbf{Background Sounds (if present)}
\begin{itemize}[leftmargin=1em, topsep=2pt, itemsep=0pt]
\item Natural (wind, rain) or Artificial (city noise, crowds). Briefly specify the environment if necessary.
\end{itemize}

\item \textbf{Music (if present)}
\begin{itemize}[leftmargin=1em, topsep=2pt, itemsep=0pt]
\item Style/genre, rhythmic features, emotional tone, atmosphere.
\item Identifiable instruments and effects (harmonies, reverb).
\end{itemize}

\item \textbf{Detailed Descriptors}
\begin{itemize}[leftmargin=1em, topsep=2pt, itemsep=0pt]
\item Changes in volume/speed/intensity. Narrative functions.
\item Detailed duration, spatial distance, pitch, timbre, texture.
\end{itemize}
\end{enumerate}

\vspace{0.8em}

\textbf{\large Important Guidelines}
\begin{enumerate}[leftmargin=*, nosep, label=\textbf{\arabic*.}]
\item \textbf{Avoid Redundancy:} Identify sources once unless they change significantly. Keep it concise.
\item \textbf{Prioritize the Audio:} Use video description \textit{only} to clarify ambiguous sounds. If a sound isn't audible, don't describe it.
\item \textbf{Avoid Hallucinated Sounds:} Only describe perceptible sounds. Avoid describing artifacts (e.g., "high-pitched squeal" from edits).
\end{enumerate}

\vspace{0.8em}

\textbf{\large Output Format}
\vspace{0.3em}

Integrate elements into \textbf{one or few sentences} following these rules:
\begin{itemize}[leftmargin=1em, nosep]
\item \textbf{Language:} English.
\item \textbf{Structure:} No lists or bullet points. 
\item \textbf{Length:} Max 40 words. Concise but detailed.
\item \textbf{Temporal Order:} Chronological (e.g., "first", "then", "suddenly").
\item \textbf{Style:} Natural, objective, context-sensitive. Focus on what is heard.
\end{itemize}

\vspace{0.8em}
\hrule
\vspace{0.5em}

\textbf{\large Examples}
\vspace{0.3em}

\textbf{Example 1 (General):}
\begin{quote}
\textbf{Input:} [High-pitched mechanical whirring with periodic thuds] \\
\textbf{Video Caption:} "Laundromat with washing machines and dryers running" \\
\textbf{Output:} Washing machines whir at high speed while dryers tumble clothes with periodic rhythmic thuds. Water drains intermittently as cycles complete and doors slam shut.
\end{quote}

\textbf{Example 2 (Anti-hallucination):}
\begin{quote}
\textbf{Input:} [Guitar strumming and melody] \\
\textbf{Video Caption:} "Musician performing with piano and guitar on stage" \\
\textbf{Output:} Acoustic guitar plays melodic fingerpicking patterns with clear, resonant tones. \textit{(Piano is omitted as it is not audible).}
\end{quote}

\end{tcolorbox}
\caption{Audio Captioning Instruction for SoundAtlas.}
\label{fig:audio_prompt}
\end{figure*}

 \end{document}